\documentclass[aps,preprint,showpacs,floatfix]{revtex4}

\usepackage{dcolumn,amsmath,amssymb}
\usepackage[utf8]{inputenc}
\usepackage[OT4]{fontenc}
\usepackage{bm}
\usepackage{latexsym}
\usepackage{color}
\usepackage{psfrag}
\usepackage{graphicx}

\def\be{\begin{equation}}
\def\ee{\end{equation}}
\def\bea{\begin{eqnarray}}
\def\eea{\end{eqnarray}}
\def\bi{\begin{itemize}}
\def\ei{\end{itemize}}
\def\bin{\begin{enumerate}}
\def\ein{\end{enumerate}}

\newcommand{\hl}{\xi}

\begin{document}

\author{M. \L{}\c{a}cki$^{a,b},$ D. Delande$^b,$ J. Zakrzewski$^{a,c}$}
\affiliation{
\ \\
$^a$Instytut Fizyki Mariana Smoluchowskiego
\\ Uniwersytet Jagiello\'nski, Reymonta 4, 30-059 Krak\'ow, Poland \\
$^b$Laboratoire Kastler-Brossel, UPMC, ENS, CNRS; 4 Place Jussieu, F-75005
Paris, France \\
$^c$Mark Kac Complex Systems Research Center
\\ Uniwersytet Jagiello\'nski, Reymonta 4, 30-059 Krak\'ow, Poland \\
}

\title{Extracting information from non adiabatic dynamics: \\ excited symmetric states of the Bose-Hubbard model}

\date{\today}
\begin{abstract}
Using Fourier transform on a time series generated by unitary evolution, we
extract many-body eigenstates of the Bose-Hubbard model corresponding to low
energy excitations, which are generated when the insulator-superfluid phase
transition is realised in a typical experiment. The analysis is conducted in a symmetric external
potential both without and with a disorder. A simple classification of
excitations in the absence disorder is provided.
The evolution is performed assuming the presence of the
parity symmetry in the system rendering many-body quantum states either symmetric or
antisymmetric. Using symmetry-breaking technique, those states are decomposed
into elementary one-particle processes.

\end{abstract}

\maketitle

\section{Introduction}
A gas of ultra-cold atoms in a sufficiently deep optical lattice is well
described by a tight-binding model --- the so called Bose-Hubbard (BH) model
as suggested by Jaksch and Zoller in their seminal article \cite{jaksch98}. The
authors predicted the occurrence of a Mott insulator to superfluid phase
transition, later realized experimentally \cite{greiner02}. Let us note
parenthetically that the notion of "Bose-Hubbard model" may be considered an
example of validity of the zeroth theorem of the history of science
\cite{jackson08}, as it should rather be called Gersch-Knollman model
\cite{gersch63}.

Optical lattices provide a superb experimental possibility not only by enabling
to implement the BH Hamiltonian but also providing means to control the
parameters of the model. Although variations of the lattice depth modifies
both the on-site interaction and the tunneling rate between sites, there is an
independent method of manipulating the strength of the interaction, by tuning
the scattering length of atoms using magnetic Feshbach resonances
\cite{court98}. Optical and microwave Feshbach resonances have also been
developed \cite{fatemi00,papoular10}.

Random on-site disorder within the Bose-Hubbard Hamiltonian causes the existence
of a new insulating, yet gap-less phase --- the Bose glass \cite{fisher89}. This
result has been generalized to the pseudo-disorder realized by a bichromatic optical
potential \cite{fallani07}.  Similarities and differences between effects due to
these two types of disorder are discussed in \cite{Roscilde08,Minguzzi08}.

To study experimentally the Bose glass phase
\cite{fallani07}, an ultra-cold atomic gas is first prepared in a trap. The optical
lattice is then switched on, driving the system through the insulator-superfluid
phase transition in a finite time. When a phase transition is classical, such a
quench may be described by the Kibble-Zurek mechanism \cite{kibble76,zurek85},
leading to the presence of several excitations in the final phase.
For quantum phase transitions, the situation was found to be
similar in an array of Josephson junctions \cite{dziarmaga02}.
Investigation of  quantum Ising models \cite{zurek05} and homogeneous BH model
\cite{polkovnikov05} shows that the number of defects scales algebraically with the
quench time.

In our previous research~\cite{zakrzewski09}, we showed that
experimental setup (approximated within the BH model) used in
\cite{fallani07} for the realization of the ground state (a series of Mott
insulators without disorder, a Bose glass with strong disorder) leads to a
significant (90\% without disorder, $>$99.999\% with the strong disorder) depletion
of the ground
state, making the interpretation of this wavepacket as a Bose glass
less obvious.

In this article, we continue our theoretical analysis by extracting eigenstates
of the Bose-Hubbard hamiltonian excited by quenching and comparing them to the
ground state and the dynamically created wavepacket. We focus on the special situation when
the external potential possesses  parity symmetry and the realization of the pseudo-disorder
(bichromatic field) respects that symmetry.

\section{The method of analysis}

As in~\cite{zakrzewski09} we take as an example the experiment of the
Florence
group (for details see the original work \cite{fallani07}). An harmonic trap was
used to confine the cold atomic gas; then a two dimensional
optical lattice potential (the ``transverse'' lattice) is ramped up to create a
two dimensional array of independent (if the lattice is sufficiently strong,
tunneling between tubes is inhibited)
one-dimensional tubes. The same ramp is used to switch on the optical potential
along the tubes. The latter potential may either be a pure ``optical lattice''
or a bichromatic lattice realizing a pseudo-random disorder.

The recoil energy $E_R=h^2/(2M\lambda^2)$ is used as the energy scale,
$\lambda=830{\rm nm}$ being the wavelength of laser beams forming the optical
lattices (both transverse and along the tube), and $M$ the mass of an atom. Initially,
the only external potential present is the
harmonic trap. Then
the 2D optical lattice creating tubes, optical lattice potential along tubes and
an additional much weaker optical lattice creating disorder are ramped together
(exponential ramping lasts 100ms in total). The additional optical lattice is
created with a different laser, with wavelength $\lambda_d=1076 {\rm nm}.$

The transverse optical lattice potential maximal height is $s_\perp=35$ (in
recoil energy  units), the maximal height of the lattice along the tubes ---
$s=14.$ Notice that we follow~\cite{zakrzewski09} in rescaling the experimental $s$ parameters
by a 7/8 factor, for a discussion see~\cite{zakrzewski09}. As it much higher than the lattice potential along the tubes,
the
transverse lattice makes the system
an independent union of 1D systems at the early stages of the ramp. We
model a single tube using the 1D Bose-Hubbard Hamiltonian (tight-binding
approximation of a full, second quantized hamiltonian):

\begin{equation}
\hat{H} = -J \sum_{j=0}^{N-1}\left(
\hat{b}_j^\dagger \hat{b}_{j+1}+ \hat{b}_{j+1}^\dagger\hat{b}_{j} \right) +
\frac{U}{2} \sum_{j} \hat{n}_{j}
\left( \hat{n}_j - 1 \right) + \sum_{j} \epsilon_j \hat{n}_j,
\end{equation}
where $\hat{b}_j$ is an annihilation  operator of one
particle at the $j$-th site. Both $J$ and $U$ depend on parameters $s$ and
$s_{\perp}$ \cite{jaksch98}. 
The lattice depth $s$
increases in time, making $U$ (slowly)
increasing and $J$ (exponentially) decreasing. The underlying
assumption within the BH hamiltonian is that the Hilbert
space is restricted only to the lowest Bloch band of the lattice. We perform the
evolution only for lattices deep enough to justify neglecting higher bands. The
$\epsilon_j$ represents the energy offset of the on-site energy at site
$j:$
 \begin{equation}
\epsilon_j=\frac{1}{2} M \omega^2 a^2 (j-j_0)^2 + s_d E_R \sin^2 \left(\frac{\pi
j\lambda}{\lambda_d}
\right).
\label{trap}
\end{equation}
The first term comes from the external harmonic trap potential ($\omega$ is the
trapping frequency), the second one corresponds to the second optical potential
introducing the disorder created by laser with wavelength $\lambda_d.$ The
parameter $s_d\ll s$ is its amplitude (in recoil energy units) --- when
$s_d=2.1875,$ it is considered strong \cite{fallani07,Roscilde08}.
As in \cite{zakrzewski09}, we use the BH model only for $s>4,$ where it is
applicable. We assume that initially the gas is in the ground state in a
superfluid state for $s=4$, and that the hopping through the transverse lattice in
negligible.

Matrix Product States \cite{mps,scholl} are used to represent the
states and the Time Evolving Block Decimation \cite{vidal03} (essentially equivalent
to Time-dependent Density Matrix Renormalization Group method  \cite{tdmrg}) algorithm is
used for the time evolution during which $s$ is increased exponentially as $s(t)
= A\left(\exp(t/\tau)-1\right).$ The exponential ramp is characterized by
$\tau=30 {\rm ms}.$ 
At the end of the ramp --- which is not adiabatic --- we have created a wavepacket,
a linear combination of the ground state and various excited
states. To characterize the properties of this dynamically created wavepacket, it is important
to know the properties of the significantly populated eigenstates.
To reach this goal, when the final $s$ value is reached, we further continue to evolve the wavepacket
using the time-independent Hamiltonian with constant $s.$ Let us
denote by {$|e_i\rangle$} the eigenbasis of the final Hamiltonian.  The
evolution of the wavepacket under constant $s$ is given by:
\be
|\psi(t)\rangle = \sum\limits_i \exp\left(-\frac{i}{\hbar} E_i t\right) c_i
|e_i\rangle,
\label{eq:psi}
\ee
and the Fourier transform (FT) of the autocorrelation function
\begin{equation}
C(t)=\langle \psi(0)|\psi(t)\rangle = \sum_i |c_i|^2 \exp \left( -i \frac{E_i
t}{\hbar} \right).
\label{eq:ct}
\end{equation}
yields $E_i$'s --- the eigenenergies of the final Hamiltonian and the overlaps
$|c_i|^2$ as discussed in \cite{zakrzewski09}. The extraction of the ground
state using imaginary time propagation was also performed.
Here, we extend this
analysis by extracting from the dynamics also the excited eigenstates with large
overlap --- those contribute most significantly to the dynamical wavepacket and
thus provide an understanding of the character of that wave function.

We focus on the situation where parity symmetry is present making the eigenstates
either symmetric or antisymmetric. This makes it
possible to apply a symmetry-breaking analysis of the final state. A short
discussion of the dynamics in a more general case without parity symmetry will
be published elsewhere.

To find a given $|e_i\rangle$ we perform a Fourier transform of the various $|\psi(t)\rangle,$
(each represented by a Matrix Product State)
on a discretized sample, $t=n\delta t,$ with $n$ an integer.
This requires a method for adding many-body states within
the MPS state representation; details concerning the method including
details on its validity will be published elsewhere.

\subsection{Bose-Hubbard model in the absence of disorder}

\begin{figure}
\begin{center}
\psfrag{Overlap}{\large{$|\mathrm{Overlap}|^2$}}
\psfrag{Energy}{\large{Energy (in recoil units)}}
\includegraphics[width=0.7\columnwidth,angle=270]{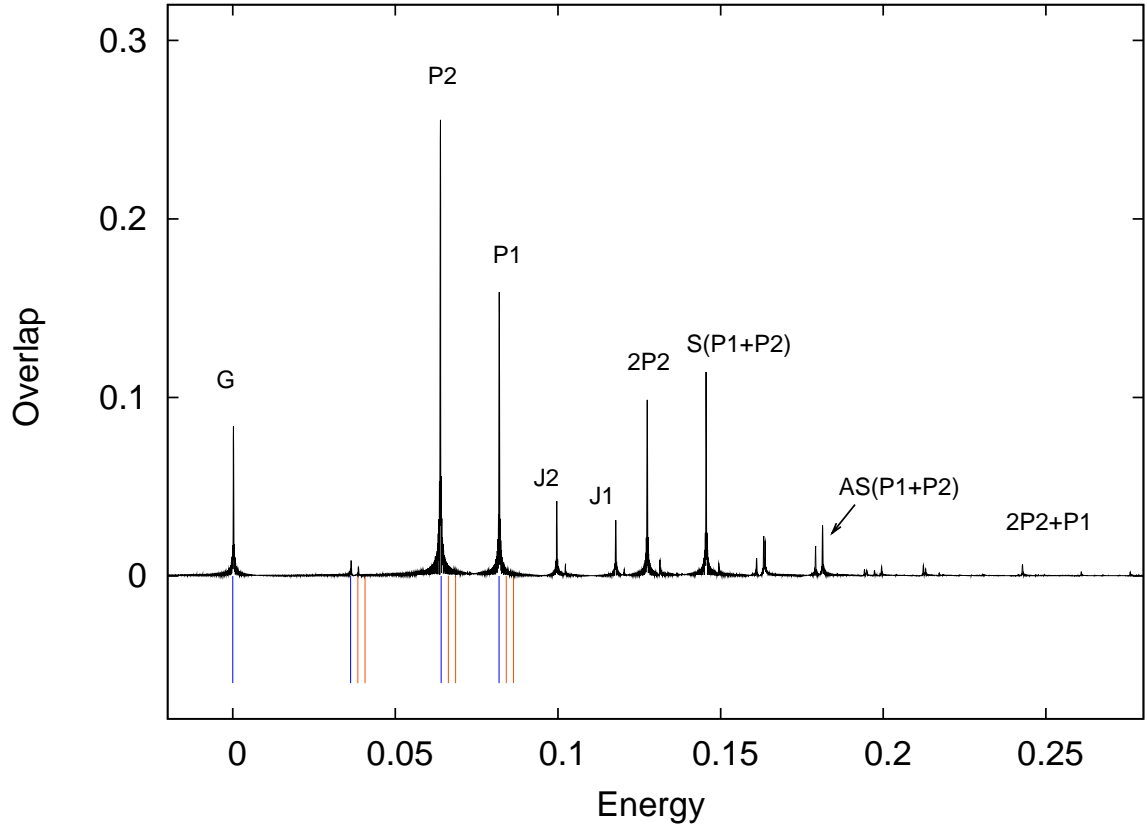}
\end{center}
\caption{(color online) Fourier transform of the autocorrelation function,
Eq.~(\ref{eq:ct}), obtained dynamically for $s_1=14$ after switching on the
lattice without the secondary lattice, i.e. for $s_d=0$ (no disorder).
The experimental \cite{fallani07} exponential ramp of 100ms is used, all other
parameters are taken as
closely as possible to the experimental situation, with $N=151$ particles on $M=81$
sites.
The peaks appear at energy levels of the system (measured with respect to its
ground state), with an intensity equal to the squared overlap with the
wavepacket.
About  ten  states are significantly excited, proving that the preparation is
not adiabatic in a strict quantum mechanical sense. The stick spectrum shown in a mirror
is the prediction for
the energies using a simple separable ansatz allowing the identification listed
in the figure --- see text for discussion.
The S(P1+P2) corresponds to a symmetric combination of both P1 and P2 excitations, the AS(P1+P2)
to an asymmetric combination. }
\label{corrd0}
\end{figure}

Consider first the example of $s_d=0,$ i.e. a pure Bose-Hubbard model.
Fig.~\ref{corrd0} shows the
FT of the autocorrelation function of the wave packet obtained using an
exponential ramp. As discussed before \cite{zakrzewski09}, the wave packet has
about 10\% (squared) overlap with the ground state (the peak at
$E\approx 119.188)$  with 4 other states contributing with higher or similar
overlap. Using the procedure sketched above
we extract the eigenstates corresponding to the dominant
contributions.

\begin{figure}
\begin{center}
\includegraphics[width=0.9\columnwidth]{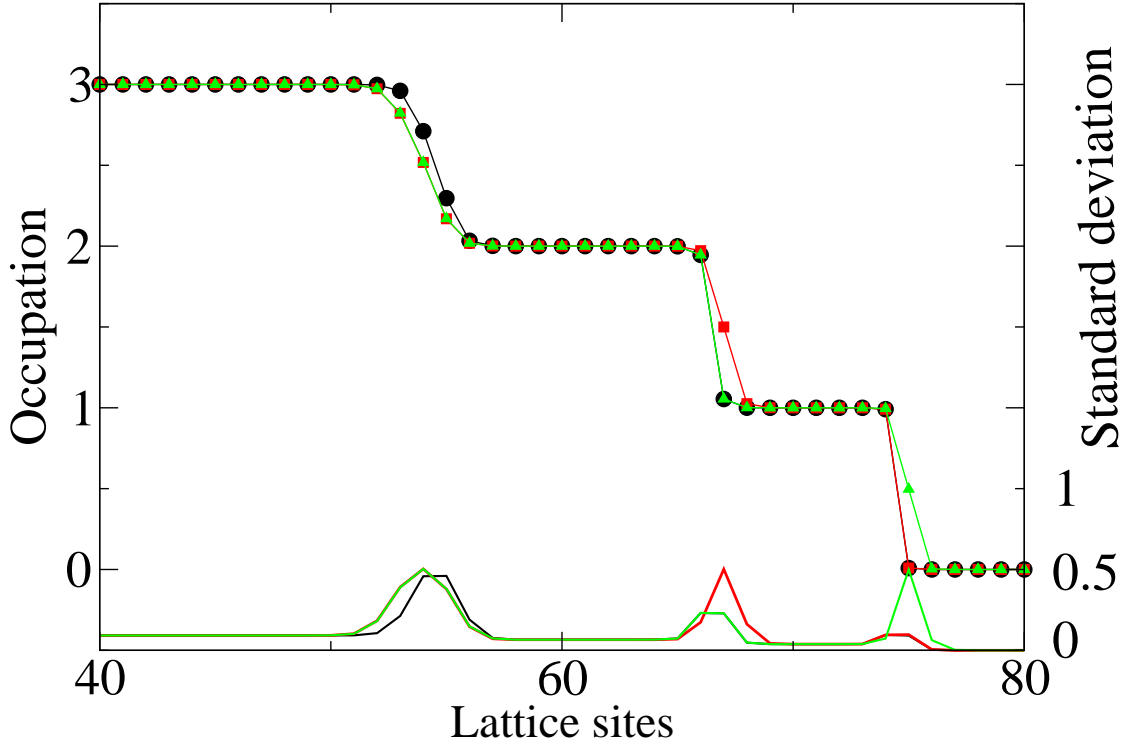}
\end{center}
\caption{(color online) Occupation of sites (left vertical axis) $\langle n_j \rangle$
for the ground
state (black circles connected by a line) and two excited states $|\psi_2\rangle$ (red)
- corresponding to the peak P2 in Fig.~\ref{corrd0} and $|\psi_1\rangle$ (green, peak
P1) which are significantly populated during the turn-on of the lattice. Due to
the symmetry of the problem, half of the system is shown only. Right horizontal
axis shows standard deviation of the occupation number
$\Delta_j=\sqrt{\langle n_j^2\rangle-\langle n_j\rangle^2}.$ Low values
correspond to
an insulating Mott state, excitations occurring in the SF zones lead to an increase of the
standard deviation.
}
\label{occup}
\end{figure}

Let us take a look at these states in some detail. In Fig.~\ref{occup}, we show
the average occupation numbers $\langle n_j \rangle$ of sites $j$ for the ground state (G) as
well as states P1 and
P2 corresponding to peaks bearing the same name in Fig.~\ref{corrd0}. Observe
that occupations of all three states coincide within broad steps of Mott plateau
(with integer
occupation of sites). For the ground state, the central $\langle n \rangle=3$
zone is broadest, for the two excited states, one particle from the
$\langle n \rangle=3$ zone is transferred either to the left or to the right.
Since we consider a symmetric potential problem (the center of the harmonic trap
coincides with  the site $j=41$), the eigenstates are either symmetric or
antisymmetric w.r.t. the trap center. This symmetry is not broken when
parameters of the Hamiltonian are varied during switching on of the lattice,
therefore only symmetric eigenstates are populated. This explains half integer
occupations on the border between $\langle n \rangle=2$ and $\langle n
\rangle=1$ zones for P2 or $\langle n \rangle=1$ and $\langle n \rangle=0$ zones
for  P1.

It seems, therefore, that basic excitations in the system correspond to transfer
of particles between edges of Mott zones. This is because
the SF regions between the zones are very small ($s=14$ corresponds to $J/U=0.0133,$ very deep in the Mott regime)
If this is the case, can the {\it most} significant excitations be explained by
such transfers ?

To test this hypothesis, let us consider first the system at $J=0$.

\subsubsection{Ground and excited states for $J=0$}
\label{sec:J=0}
For $J=0,$ the ground state is well known. All eigenstates, in
particular the ground state, are product of Fock states at different sites
--- at each site $i$ we put exactly $n_i$ particles.
\be
|\psi\rangle = \bigotimes\limits_{i=0}^M | n_i\rangle
\ee

\begin{figure}[ht]
\centering
\psfrag{kx}{$k\hl$}
\psfrag{gx}{$\gamma\hl$}
\includegraphics*[width=0.6\linewidth,angle=0]{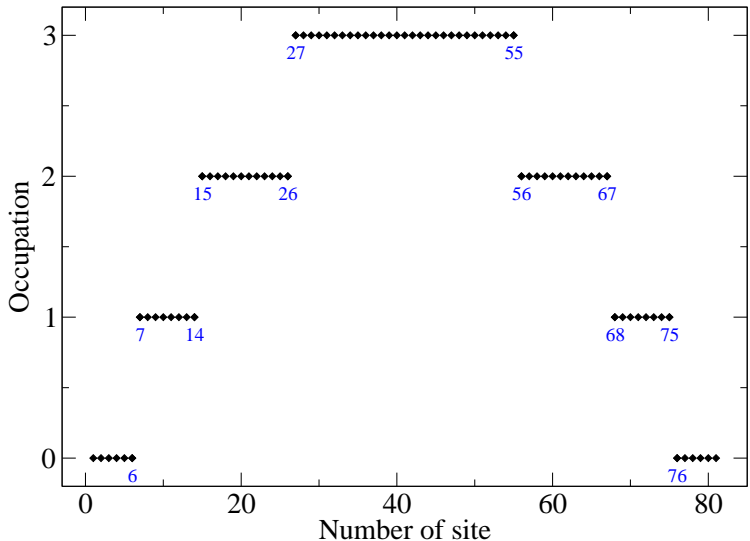}
\caption{Ground state of the BH model for $J=0,$ composed of a series of insulating Mott plateaus
with the familiar wedding cake shape. In blue, sites at the
edge of the Mott zones.}
\label{fig:groundJ0}
\end{figure}

To create the ground state with $N$ particles, one has to perform $N$ time the
following procedure: find the site whose local energy (energy
taking into account the harmonic trap (\ref{trap}) together with interaction
with particles already present on the site) is the least and put the particle into that site.
Without disorder,
the ground state has a well-known wedding cake form - see
Fig.~\ref{fig:groundJ0}.

\begin{figure}[ht!]
\centering
\psfrag{kx}{$k\hl$}
\psfrag{gx}{$\gamma\hl$}
\includegraphics*[width=0.7\linewidth,angle=0]{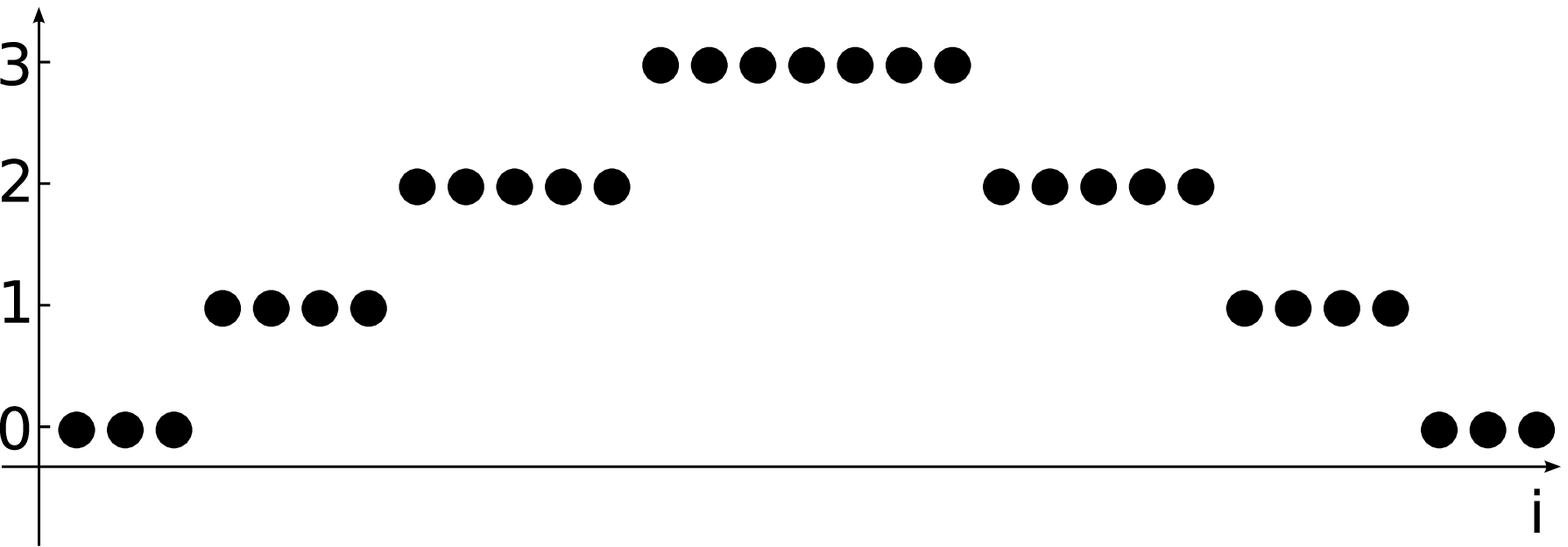}
\includegraphics*[width=0.7\linewidth,angle=0]{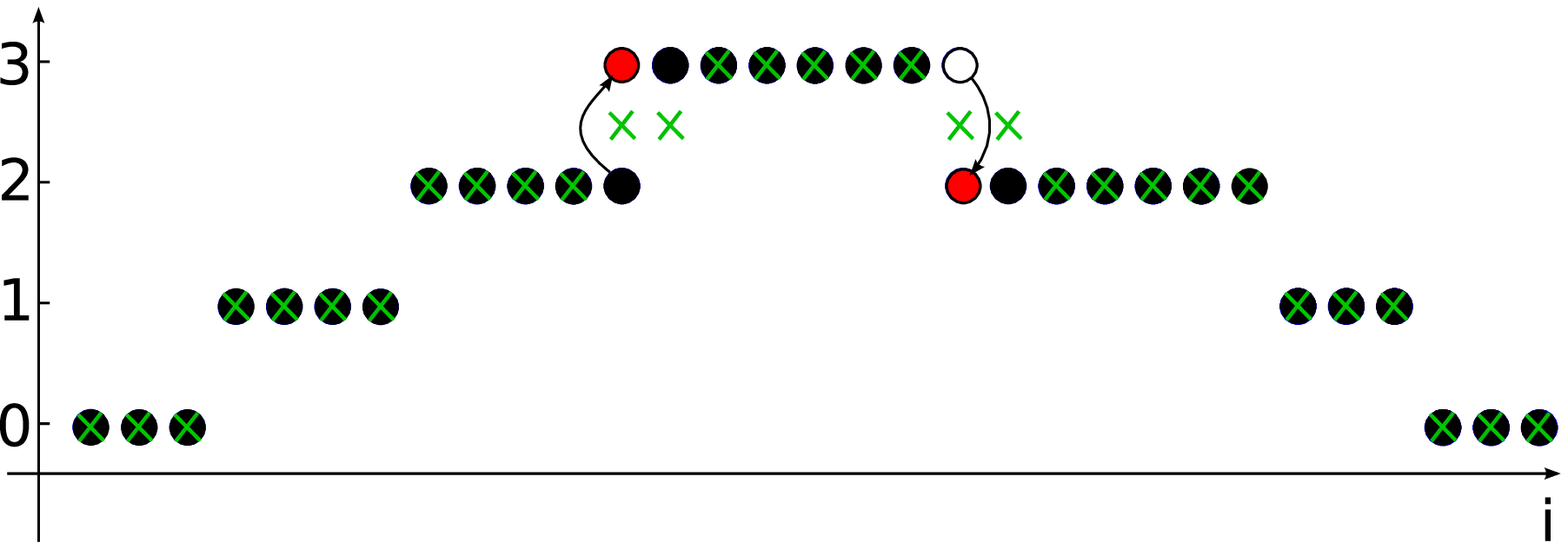}
\includegraphics*[width=0.7\linewidth,angle=0]{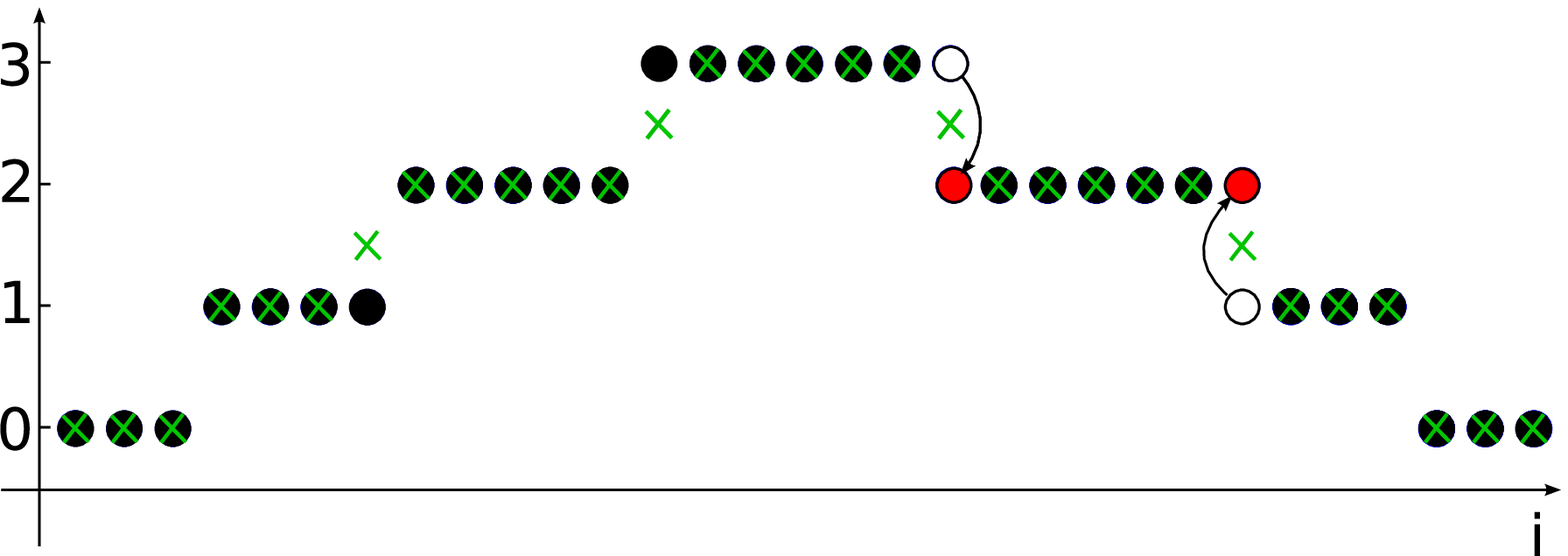}
\includegraphics*[width=0.7\linewidth,angle=0]{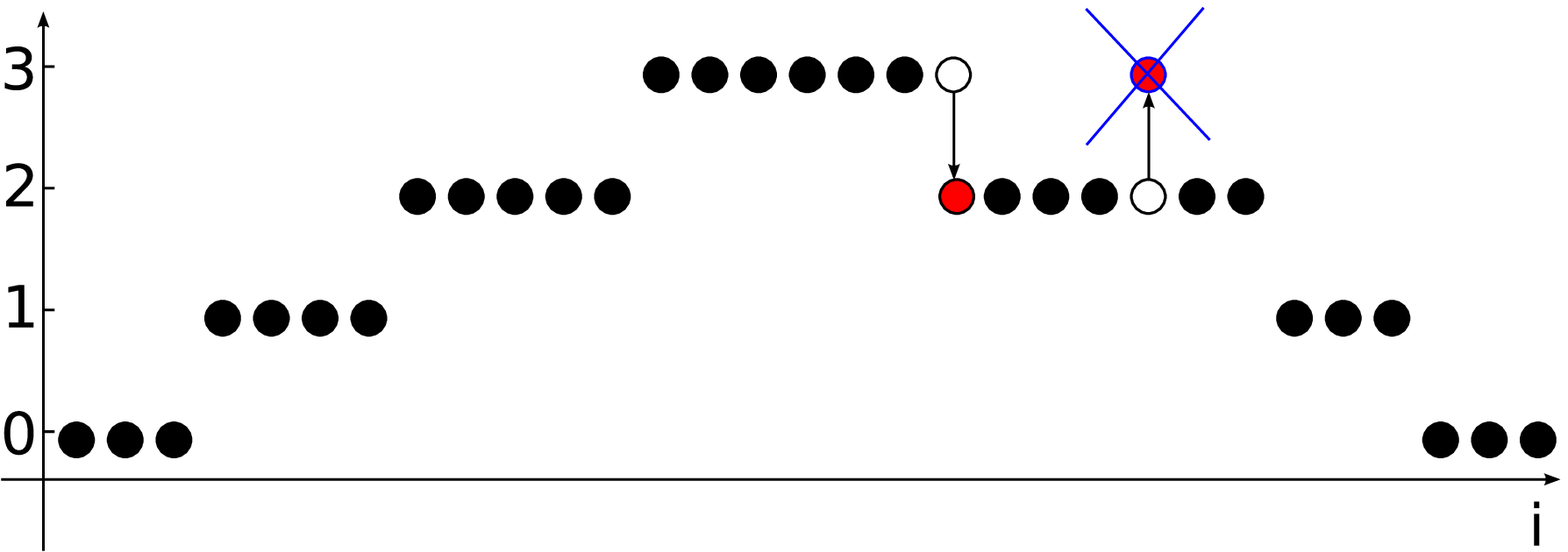}
\caption{(color online) (a) Ground state of the Hamiltonian for $J=0.$ For a
better visualization, the number of sites is reduced from $M=81$ in simulations
of the experiment to $M=35$. Particle removed from a certain MI zone may be put
back at the same zone, but on the other side from the center of the trap
effectively shifting the zone to one side (b), or to other level (c). After
symmetrization of the state, a distribution denoted by green crosses will be
obtained. Bottom plot shows an example of other possibility that exists for
$J=0.$  For $J>0,$ this type of excitation is unlikely to be
observed in a real experiment,
due to its dynamical instability and a too high excitation energy.}
\label{fig:cake}
\end{figure}

To get the a low-lying excited state, one should take a few particles (perhaps
one) and move it to a different place. To achieve low energy difference, the
particle must be put back in such a way that the resulting particle distribution
is close to optimal. Some possibilities are shown in Fig.~\ref{fig:cake}.

Consider now our exemplary case of $N$=151 particles on 81 sites. The ground
state at $J=0$ is shown
in Fig.~\ref{fig:groundJ0}. At site numbers marked in the figure, there are
sudden changes of local occupation numbers defining the edges of different occupation
zones (Mott plateau). As the ground state is $\mathbb{Z}_2$-symmetric (i.e. even
with respect to reflection at the center of the trap), several
$\mathbb{Z}_2$-symmetric one-particle excitations may be generated by taking one
particle from one of sites 7, 15 or 27 and moving it to one of sites 6, 14 or 26.
Symmetric possibilities are when the particle is taken on the right side
(sites 75, 67, 55 symmetric partners of respectively 7, 15, 27) and/or
is moved to sites 76, 68, 56 on the right side.
This generates 9 different energy combinations that are listed in Table \ref{tab:1pexc}, each combination
being 4 times degenerate (left/right for the particle removed, left/right
for the particle added).

\begin{table}
  \caption{\ \\
Comparison of 9 elementary excitations for $J=0.$ Each one particle
excitation corresponds to moving a particle from site $i$ to site $j$ (due to symmetry w.r.t site 41, only
excitations where a particle is moved from the left side to the right side are shown).
Third column gives the energy of that excitation. Fourth column gives the excess
energy with respect to the ground state. Due to symmetry w.r.t site 41, only
excitations where a particle is moved from the left side to the right side are shown. Fifth column identifies
the state with state obtained for small $J$ in Fig. \ref{corrd0}. Last column
give the degeneracy of excitation for $J=0$, with, inside the parentheses,
the degeneracy expected in the low $J$ limit.}\vspace{2mm}
  \begin{tabular}{||l|l|l|l|l|l||}
  \hline
  $i$ & $j$ & Energy & Excess energy & Notation & Degeneracy\\
  \hline
  \hline
   - & - & 119.332 & 0 & G  &1\\
27&56&119.368&0.036&S3&4(2)\\
15&56&119.370&0.038&T2&4\\
 7&56&119.373&0.041&T1&4\\
27&68&119.396&0.064&P2&4\\
15&68&119.398&0.066&S2&4(2)\\
7&68&119.401&0.069&D1&4\\
27&76&119.414&0.082&P1&4\\
15&76&119.416&0.084&D2&4\\
7&76&119.418&0.086&S1&4(2)\\
\hline
\end{tabular}

\label{tab:1pexc}
\end{table}

\subsubsection{Nonzero small $J$, no disorder}

Let us come back to the ``physical'' case of small $J/U=0.0133$ corresponding to
the lattice with height $s=14$ and the FT of the correlation function of
dynamically obtained wave packet shown in Fig.~\ref{corrd0}.
The ground state energy $E=119.185$ is quite close to $J=0$ estimate
(see Table \ref{tab:1pexc}) - the difference is on the fourth significant
digit. Still this difference is much larger than the excess energy of
excitations listed in the Table - it is mainly due to the important role of
tunneling in the narrow SF strips. The exemplary extracted states, see
Fig.~\ref{occup}, suggest
that the excitations are due to transfers between edges of Mott zones. Therefore,
the ``excess energy'' of such excitations above the ground state can be
approximated by the $J=0$ energy excess (remember we consider deep optical lattice -
this argument might not hold close to the SF-MI transition). By inspection of
the excess energies, we can then identify {\it several} important contributions
to the wave packet in terms of elementary excitations and their multiplicities!
The classification is included in Fig.~\ref{corrd0}.

It is apparent that the main origin of the nonadiabaticity comes from P1 and P2
excitations and their multiplicities. The corresponding elementary particle
exchange process is a loss of particle from  the highest $\langle n \rangle=3$
Mott zone  to $\langle n \rangle=2$
(P2) or  to $\langle n=1 \rangle$ (P1).

Excitation of type S are slightly different: indeed, there, the particle
is removed from Mott plateau $n$ and added in Mott plateau $n-1.$ When
the two sites are on opposite sides, it simply corresponds to a right or left
shift of plateau $n$ by one site. For example, the S3 component has a squared overlap of
less than 1\%, still the analysis of the associated state
by our method confirms that assignment. This process for other Mott zones (S2 and S1) is not observed.
There is however another possibility for S excitations: if the $i$ and $j$ neighboring sites are on the same side,
it is associated with a hole in the $n$-plateau (can be viewed also as an extra particle
in the $n-1$ plateau), and --- because $i$ and $j$ are neighbors --- it costs some kinetic energy.
Thus one can expect the 4-fold degeneracy of the S excitations at $J=0$ to be rapidly
lifted to a 2-fold degeneracy only for relatively small $J$. Also, the same-side excitation
is dynamically unstable and unlikely to be significantly excited in our wavepacket.

We do not observe significant contributions of processes involving
exchange of particles between $\langle n \rangle=1$ to $\langle n \rangle=2$
zones only (D type). A qualitative explanation for that fact might be that
formation of $\langle n \rangle=1$ and $\langle n \rangle=2$ zones takes time
earlier in the ramping-up process (for larger $J/U$). Then tunneling more
efficiently redistributes particles between sites, also the exponential ramp
changes relatively slower. Moreover, observe that excitations when the particle
is promoted to the highest $\langle n \rangle=3$ level at the expense of
shrinking the lower Mott zone (T1 and T2) are not observed. This seems
physically understood quite naturally. With the increase of the lattice depth
the highest Mott zone also increases at consecutive avoided crossings if passed
adiabatically. If it does not have time to increase sufficiently, P1 or P2
excitations are created. But no avoided crossings of T1 or T2 process may
occur.

\begin{figure}[ht]
\centering
%\psfrag{kx}{$k\hl$}
%\psfrag{gx}{$\gamma\hl$}
\includegraphics*[width=0.9\columnwidth]{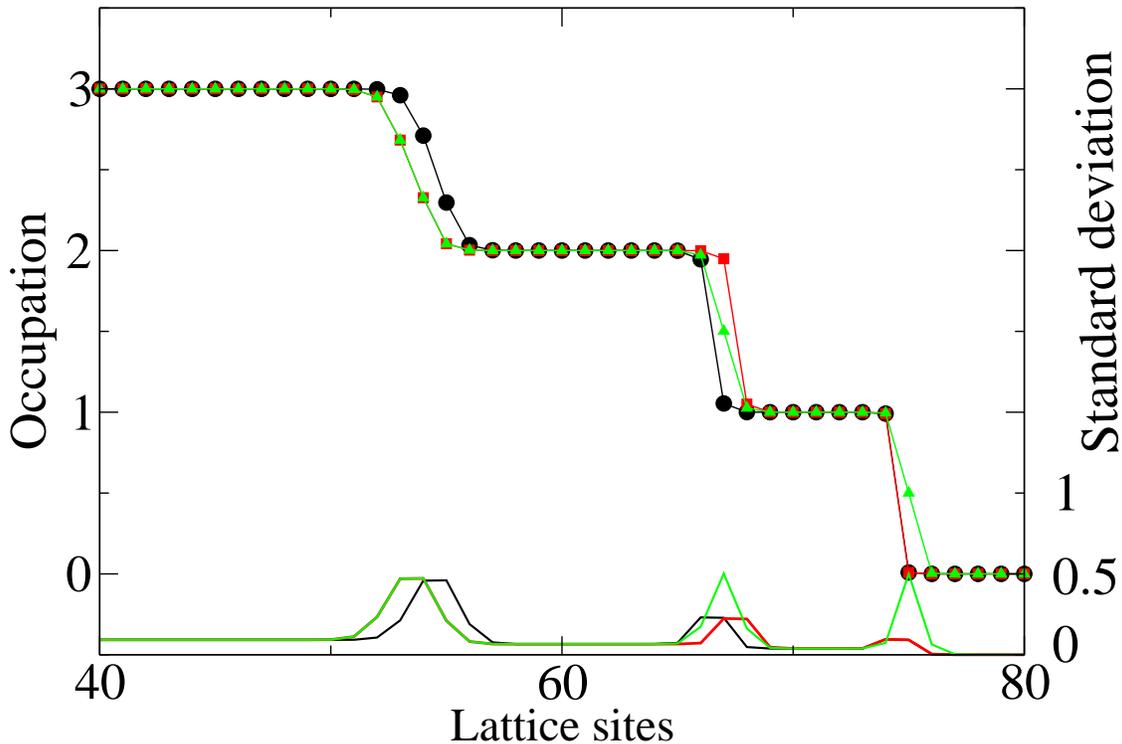}
\caption{(color online) Exemplary excitations involving transfer of two atoms
from the $\langle n \rangle =3$ zone, either both to $\langle n \rangle =2$ (one on
the left one on the right) - 2P2 excitation, red line; or one to $\langle n
\rangle =2$ and the other one to $\langle n \rangle =1$ - P1+P2 (green). Due to the
symmetry of the potential only half of the trap is shown. The right vertical
axis shows standard deviation of the occupation number.
}
\label{fig:multi}
\end{figure}

\begin{figure}[ht]
\centering
%\psfrag{kx}{$k\hl$}
%\psfrag{gx}{$\gamma\hl$}
\includegraphics*[width=0.9\columnwidth]{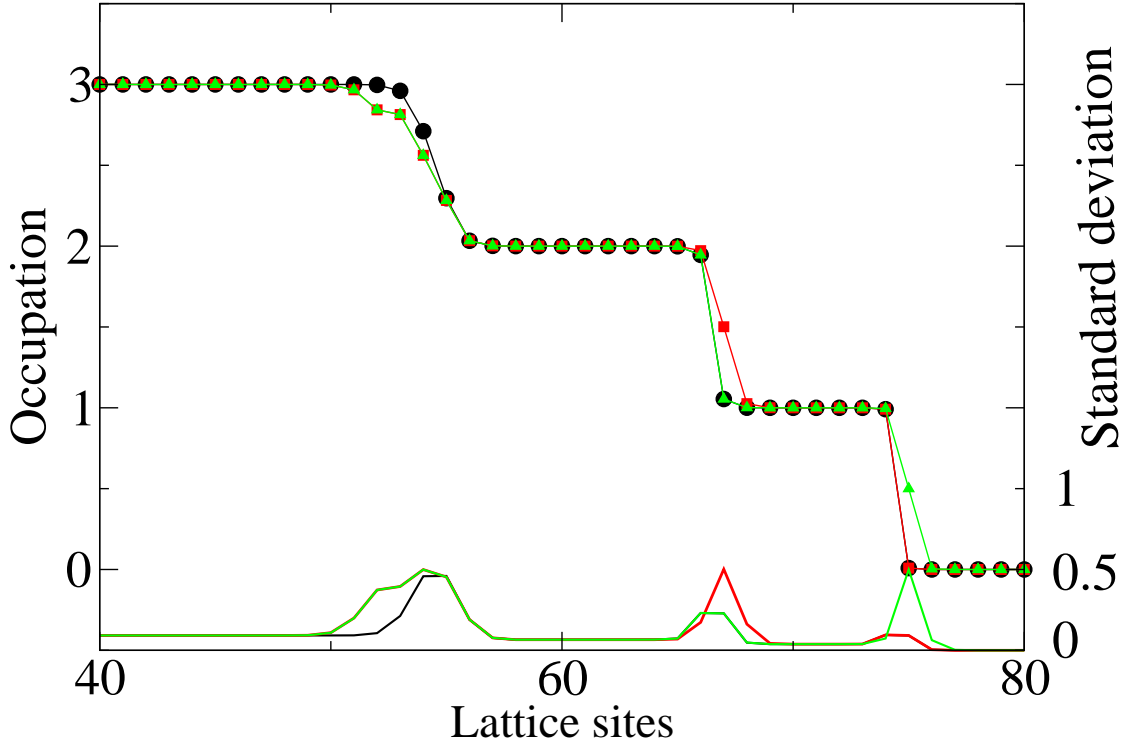}
\caption{(color online) Examples of excitations involving particles in
superfluid zones.
Excitations denoted as J2 and J1 in Fig.~\ref{corrd0} are shown as red and green
symbols. The excitation leads to broadening of the SF zone separating the $\langle n
\rangle =3$ and  $\langle n \rangle =2$ Mott zones. Due to the symmetry only
half of the trap is shown. The right vertical axis shows standard deviation of
the occupation number. The J2 excitation corresponds to the transfer of one particle
from a SF zone to a $\langle n \rangle =2$ Mott plateau, J1 to the $\langle n \rangle
=1$ zone as could be traced back from the standard deviation plot. The standard
deviation of the occupation number (right scale) confirms the broadening of the
SF region between the $\langle n \rangle =3$ and $\langle n \rangle =2$ Mott zones.}
\label{fig:jexcit}
\end{figure}

With that in mind, we can actually identify the dominant processes contributing
to the P1 and P2 excitations.
Consider P1 only (for P2 the same arguments apply except different lower Mott
zones are involved). Recall that P1 corresponds to loss of one particle at the edge
of the $\langle n\rangle=3$ zone, say on the left side (site $i=27$) with
the additional particle appearing at the edge between the $\langle n\rangle =0$ and
$\langle n \rangle=1$ zones. This may occur at the same side of the wedding cake
(site $j=6$ in our example) or on the opposite side of the center (at site
$j=76$). A more detailed analysis is given in the next section.

Similarly, we may easily understand the P1+P2 or 2P2 processes
(compare Fig.~\ref{fig:multi}) contributing significantly to the FT of the
autocorrelation function shown in Fig.~\ref{corrd0}. Then, the Mott $\langle n
\rangle=3$ zone loses one particle at each side, both particles moving to
lower zones.
The 2P2 case is symmetric, on both sides a particle appears at the border
between $\langle n \rangle =1$ and $\langle n \rangle=2$ zones.

For P1+P2 a symmetric combination is created, P1 on the left and P2
on the right or vice versa. The classification denoted in the correlation function plot have been confirmed
by extracting the excited states responsible for those peaks.

Apart from simple excitations, easily identifiable in the $J=0$ limit, there are
other excitations involving particles in narrow superfluid strips separating
different Mott zones. Two of these excitations, denoted as J1 and J2 in
Fig.~\ref{corrd0} are shown in Fig.~\ref{fig:jexcit}.

\section{Parity Symmetry breaking}

It is possible to describe the state P1 (this reasoning is not limited to this
state, the presented method is rather general) further by describing
correlations within eigenstates of $\mathcal{H}.$
As shown in section~\ref{sec:J=0}, in the limit $J\to 0,$ the
P1 excited state has a 4-fold degeneracy, with 2 states symmetric
by parity symmetry and 2 antisymmetric ones. When both the Hamiltonian
and the dynamically excited wavepacket is symmetric, the overlaps with
antisymmetric states vanishes, making them invisible in the FT of the autocorrelation
function. The two remaining (symmetric) states are strictly degenerate only for $J=0.$
For non-zero $J,$ they are coupled via tunneling of one particle from one side of the $\langle n \rangle=3$ plateau
to the other side. The associated amplitude is very small, and the two
states are almost degenerate, that is not resolved in out FT over a finite time
interval.

However, if we now break the symmetry parity and propagate the same wavepacket with
a slightly asymmetric Hamiltonian --- for example obtained by shifting the trap center by $\delta\ll 1$
with respect to the lattice --- the 4-fold degeneracy as well as the selection rules forbidding
the excitation of the antisymmetric states will be broken, and one expects to observe the P1 peak to
split in a multiplet of 4 peaks.
\begin{figure}
\psfrag{Overlap}{$|$Overlap$|^2$}
\psfrag{Energy}{Energy}
\includegraphics[width=12cm]{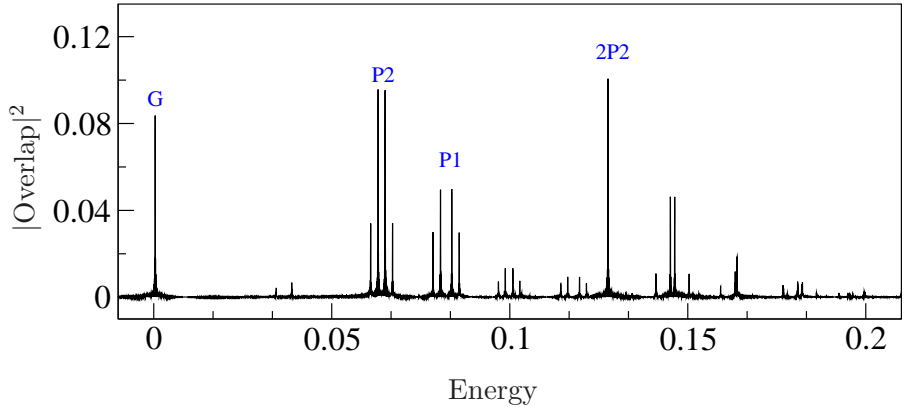}

\caption
{
Fourier Transform of the autocorrelation of a wavepacket evolved for
$\delta=0.03.$ One clearly observes splitting of eigenenergy peaks with a
multiplicity corresponding  to the degeneracy at $\delta=0$ given in table~\ref{tab:1pexc}.
 Labels  associate the 4-fold group of peaks for the symmetry broken
case  to a corresponding single peak at $\delta=0$. Note that the 2P2 peak does not split, as it corresponds to a non-degenerate excitation.
}
\label{fig:widelec}
\end{figure}

The FT of the
autocorrelation function in Fig. \ref{fig:widelec} clearly shows these 4 components.
As expected the height of the P1 peak for $\delta=0$  --- 0.158 --- is now shared by the four
components of the multiplet, the sum of heights being 0.154.

For $J=0,$ each of constituent of the P1 multiplet differ from
the ground state by a particle jump from one of sites 27 and 55 to one of
sites 6 and 76. These particle jumps break the parity symmetry and make the
energies dependent on $\delta$ in the first power.  Outer peaks
correspond to jumping particles over a long distance - 55 to 6 or 27 to 76,
inner, higher peaks correspond to one sided jumps: 27 to 6 and 55 to 76.
It turns out that values $E_0^i-E_\delta^i$ calculated for $J=0$ are very close
to those extracted from Fig.~\ref{fig:widelec}. 

It might have been tempting to argue that the larger the distance between two
sites $i$ and $j$ the lesser the probability of finding an excitation that
differs from the ground state by hopping a particle from site $j$ to site $i.$
The heights of peaks that are results of splitting an symmetric ground state
show a different picture. The probability is certainly lower, but the order of
magnitude remains largely the same --- the difference is by a factor of 1.5-4
depending on the symmetric eigenstate being considered.

The P2 peak behaves exactly like the P1 peak with a quadruplet appearing when parity is broken, see fig.~\ref{fig:widelec}.
The 2P2 peak behaves very differently, with a single peak surviving keeping all the weight. As noticed above,
the 2P2 case corresponds to 2 atoms jumping from the $\langle n\rangle=3$ to each of the  $\langle n\rangle=2,$
a symmetric non-degenerate state, like the ground state, in the $J=0$ limit.

\begin{figure}
\begin{center}
\psfrag{Overlap}{\large{$|\mathrm{Overlap}|^2$}}
\psfrag{Energy (in recoil units)}{\large{Energy (in recoil units)}}
\includegraphics[width=0.7\columnwidth,angle=270]{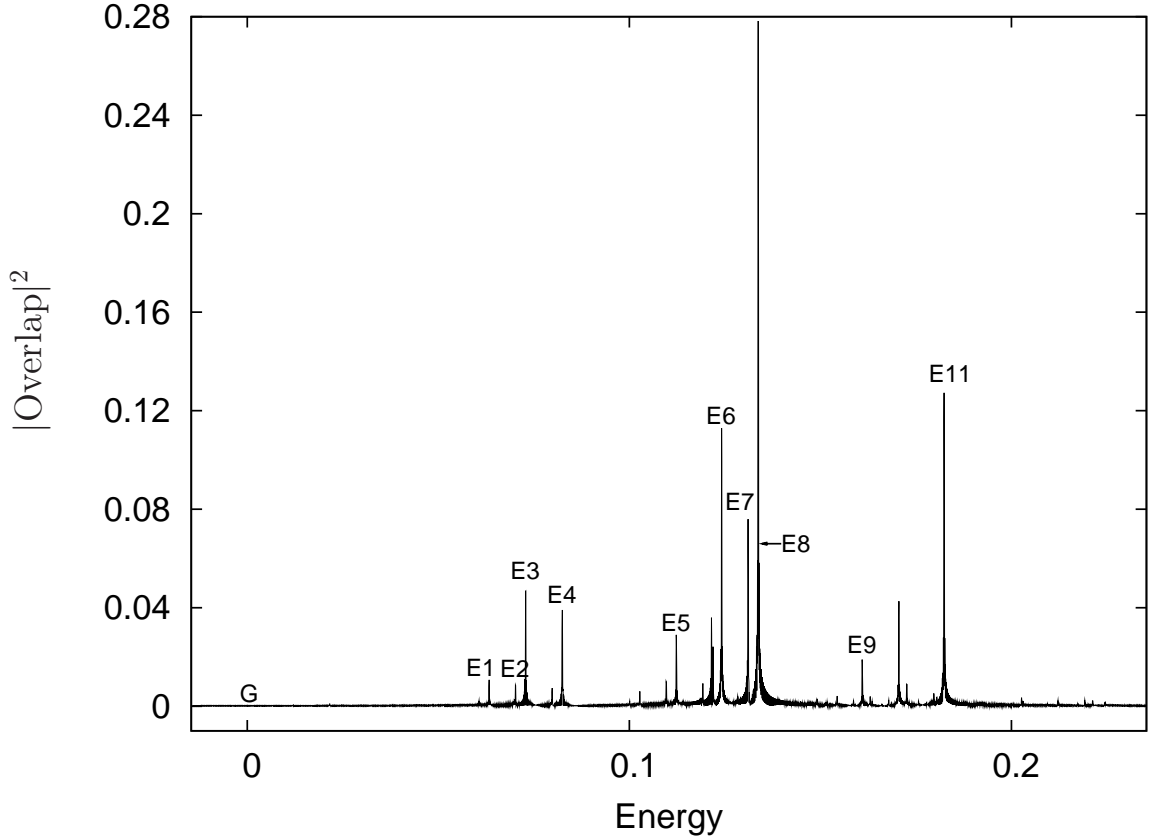}
\end{center}
\caption{(color online) Fourier transform of the autocorrelation function,
Eq.~(\ref{eq:ct}), obtained dynamically for $s_1=14$ after switching on the
lattice with the secondary lattice of strength $s_d=2.1875$ (151 particles on 81 sites).
Several peaks represent excitations populated during turning on of the lattices.
The ground state is populated with less than 1ppm probability.}
\label{corrd25}
\end{figure}

\section{The disordered BH model}

The presence of the on-site disorder, e.g. in the form (\ref{trap}) for $s_d\ne
0 $ modifies both the dynamics and the static properties of the BH model. An exhaustive
analysis of possible phases has been published in~\cite{Roscilde08,Minguzzi08}.
In our system the disorder is due to
bichromaticity of the lattice, created by using two incommensurate laser pulses.
This is not entirely equivalent to true random on-site potential as in
\cite{fisher89}. Nevertheless for such a strong disorder, for not too big $J$
the system is in a gap-less, insulating Bose glass phase.

\begin{figure}
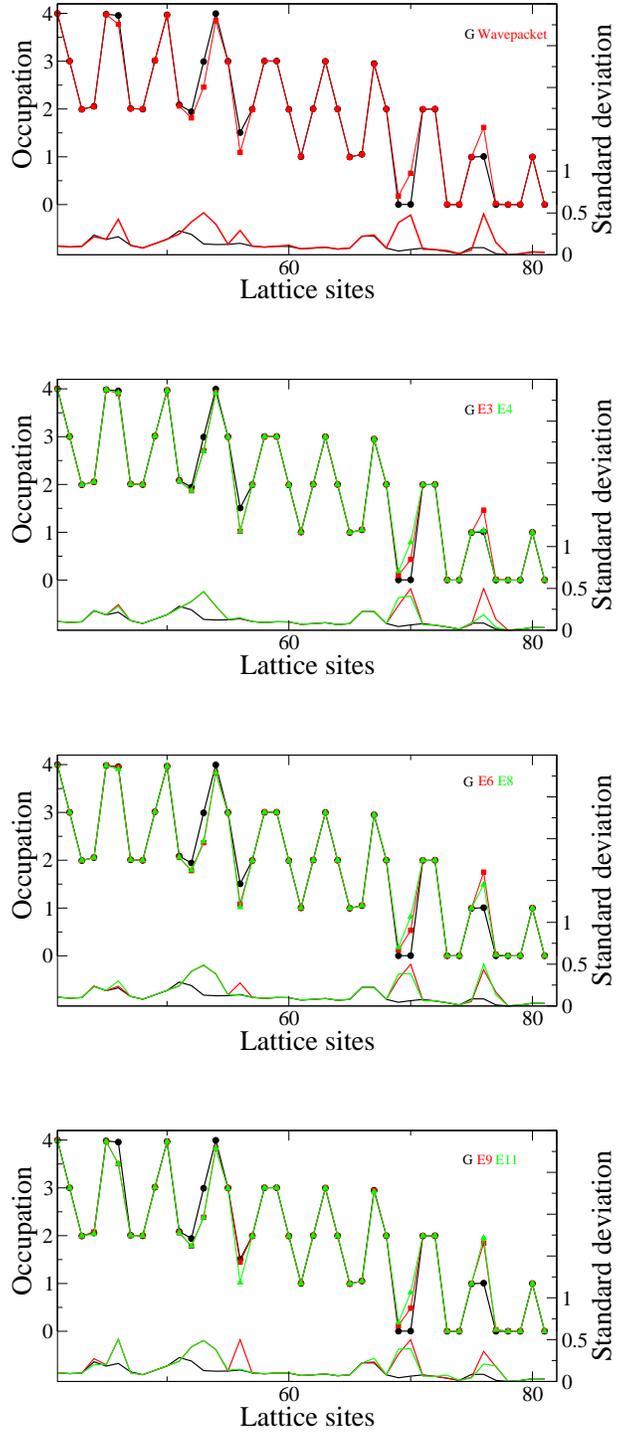

\begin{center}
\includegraphics[width=8cm]{ocGVP.eps}
\vskip 1truecm

\includegraphics[width=8cm]{ocE3E4d25.eps}
\vskip 1truecm

\includegraphics[width=8cm]{ocE6E8d25.eps}
\vskip 1truecm

\includegraphics[width=8cm]{ocE9E11d25.eps}
\end{center}
\caption{(color online) The top graph compares the dynamical wavepacket and
the ground state, subsequent graphs compare the ground state (black on each
plot) to the significantly populated excited states constituting the wave packet
obtained for $s_d=2.1875$. The right hand axis shows standard variation of the
occupation number.}
\label{fig:example25}
\end{figure}

Note that the spectrum of the autocorrelation function, Fig. \ref{corrd25}, is
indeed much denser than in the absence of disorder: this is easy to understand,
as random variance of the local energy increases likelihood that several sites
have similar on-site energies, providing opportunity for low energy
excitations.  For $J=0,$ the analysis in terms of
local energy cost for adding a new particle, developed in the absence of disorder,
is still valid. The difficulty is that --- because of the random fluctuations of the on-site energy ---
there is no way of classifying the sites as attached to a well defined plateau.
This implies that a slightly different realization of the disorder
will produce a different landscape of low energy excitations.
Nevertheless, only a handful of states are
significantly excited. In Fig. \ref{fig:example25}, the most significant excitations
have been plotted. A noteworthy difference with the no disorder case is that,
although excitations are local in nature, they are not single particle
excitations. Each excited state and the ground state have similar occupation and
particle number variance at all sites, except a few sites where the occupation
is different, and particle number fluctuations are bigger.

The analysis performed in the previous section may also be done for the
symmetric eigenstates. The results are analogous --- peaks in the
autocorrelation function graph split into several peaks corresponding to
symmetry-broken eigenstates of the unsymmetric system.

\section{Conclusions}

We have described how the excited states may arise when evolving an
ultracold gas in an optical potential --- as a result of going through
a superfluid-insulator quantum phase transition. In the absence of disorder, the
excitations are local,  one or a few particles are misplaced with respect to the
ground state. The excitations appear only between edges of the Mott regions
creating long range correlations.

The wavepacket is a sum of states differing from the ground state by a single
or a few particle elementary excitations and therefore similar in nature. Most
notably in all presented states (in the Mott insulator region), all states have almost
the same average occupation, very similar particle number variance. Still the
wavepacket is not a single eigenstate but a quantum superposition. This fact
leads to the rise of nonlocal correlations throughout the sample. It may be
directly observable as an increased variance of nonlocal observables such as number
of particles in one half of the system. If one measured the number or particles
to the left of the middle site, the statistical distribution stemming
from independent realizations of the experiment would have a large number
variance, much larger than the local particle number variance at the given site.
This directly follows from the symmetry-breaking description and the presented data, as
different eigenstates building the wavepacket have different number of particles
in one half of the system (jump of a particle from one half of the system
to the other) than in the other half.

Slight symmetry breaking allows to analyze constituents of the symmetric
eigenstates. It turns out that excitations differing by a long range particle
jump are less probable, but not completely negligible.

A few questions arise. Firstly, we have considered only the deep lattice
regime. If the lattice was shallower, what would be the nature of excitations
present in the system? Secondly many more excitations would be present if the
system was in the significantly nonzero temperature - how would it affect
properties of the system such as the mentioned long range coherence and its
decoherence?

\section{Acknowledgments}

We are grateful to K. Byczuk for pointing out the Gersch and Knollman paper
\cite{gersch63}.
Support within Polish Government scientific funds for 2009-2012
as a research project is acknowledged. M. \L. is  grateful for
support within Jagiellonian University International Ph.D
Studies in Physics of Complex Systems (Agreement No.
MPD/2009/6) provided by Foundation for
 Polish Science and cofinanced by the European
  Regional Development Fund. Computer simulations were performed at
  ACK Cyfronet AGH as a part of the POIG PL-Grid project (M\L) and at ICM UW
under Grant No. G29-10 (JZ and M\L).

\end{document}